\documentstyle{article}
\title{
\begin{flushright}
{\bf\normalsize   LPTHE-Orsay-94-59}\\ \end{flushright}
\bf A Remark on the Renormalization Group Equation \\
for the Penner Model}

\author{ {\it D.A. Johnston}\\
         LPTHE\\
	 Universite Paris Sud, Batiment 211\\
         F-91405 Orsay, France$^{1}$\\
	 \\
         Dept. of Mathematics\\
         Heriot-Watt University\\
         Edinburgh, EH14 4AS, Scotland$^{2}$ }

\begin{document} \maketitle
		      {\Large \begin{abstract}
%
It has been shown that it is possible to extract values
for critical couplings and $\gamma_{string}$ in matrix models
by deriving a renormalization group equation for the variation
of the of the free energy as the size $N$ of the matrices
in the theory is varied.
In this paper we derive a ``renormalization group equation''
for the Penner model by direct differentiation
of the partition function and show that it reproduces
the correct values of the critical coupling and $\gamma_{string}$
and is consistent with the logarithmic corrections present for $g=0,1$.
\\ \\
\\ \\ \\ \\ \\ \\ \\ \\
$1$ {\it Address Sept. 1993 - 1994} \\
$2$ {\it Permanent Address}
%
			\end{abstract} }
%
  \thispagestyle{empty}
%
%
  \newpage
%
		  \pagenumbering{arabic}
It was observed by Brezin and Zinn-Justin \cite{1} that the
known double-scaling behaviour of
the free energy $F$ of the matrix models associated with
$c \le 1$ theories
\begin{equation}
F \simeq \Delta^{2 - \gamma_0} f ( \Delta N^{2 / \gamma_1} )
\end{equation}
where $\Delta = g - g_c$, $\gamma_0 + \gamma_1 = 2$
and
$\gamma_1 = (1 / 12) ( 25 - c + \sqrt{(1-c)(25-c)} )$
meant that they should satisfy a renormalization group
equation of the form
\begin{equation}
\left( N {\partial \over \partial N} - \beta (g) {\partial \over \partial g} +
\gamma (g)
\right) F = r(g)
\end{equation}
where $\gamma (g) = 2$,
with the critical coupling(s) determined by $\beta(g_c) = 0$ and
the scaling exponents by $\gamma_1 = 2 / \beta'
(g_c)$. Such an approach, considered before in other contexts \cite{2},
is of interest as a possible means of extending matrix model calculations
beyond the so-called $c=1$ barrier.
They showed that a lowest order calculation,
in which a row and a column were explicitly integrated out in
a matrix model at finite $N$, gave values that
were reasonable and suggested that pushing the calculation further would
improve the numerical accuracy. However, a direct continuation of this approach
proved disappointing, with no sign of a convergence to the known
critical couplings \cite{3}.
Similarly, applying the methods to a $d=1$ matrix model
gave results that were correct qualitatively
but not very accurate quantitatively \cite{3a}.
The vital missing insight came from
work on the renormalization group for vector models \cite{4}, where
it was observed that it was necessary to take account of the
reparameterization invariance of the theory via the Schwinger-Dyson (ie loop)
equations in order to eliminate unwanted induced couplings.
The same approach gave excellent results when applied to $d=0$ matrix models
as well \cite{5}, whether in a eigenvalue representation or by direct
integration
of a row and column in the style of Brezin and Zinn-Justin.

If we integrate out a single eigenvalue $\lambda$ in the style of \cite{5} to
take us from a $N+1 \times N+1$ matrix model to a $N \times N$ matrix model
then we
find, symbolically, a modified action
\begin{equation}
Z_N \simeq \int d \lambda \int d^{N^2} \Phi  \exp \left[ - (N+1)  tr V(\Phi)
- (N+1) V(\lambda)
 + 2 tr \log | \lambda - \Phi| \right]
\label{e01}
\end{equation}
where $V(\Phi)$ is the original action.
This in turn leads to a renormalization group equation
as $N \rightarrow \infty$ for the
free energy $F$
\begin{equation}
\left( N {\partial \over \partial N} + 2 \right) F = { 1 \over N}
 <tr V ( \Phi) >   + V ( \tilde
\lambda ) - { 2 \over N } < tr \log | \tilde \lambda - \Phi |>  + ......
\end{equation}
where $\tilde \lambda$ is the saddle point value of $\lambda$ arising from
carrying out
the integration in equ(\ref{e01}). The use of the loop equations,
which encode the reparameterization invariance of the theory, enables one to
eliminate the complicated induced interactions arising from the logarithmic
term
and get back to a beta function that depends only on the original couplings
\begin{equation}
\left( N {\partial \over \partial N} + 2 \right) F = G \left( g , {\partial F
\over \partial
g} \right) + ......
\end{equation}
at the expense of including terms nonlinear in $\partial F / \partial g$ in
$G$.
The difficulties arising in the application of a renormalization group
procedure
to the $d=0$ matrix model reside essentially in the induced logarithmic term
that is produced by the integration over an eigenvalue (or, equally, a row and
a
column). It is thus not unreasonable to ask what happens in the case where such
a term
is already present in the original action. Such models have been considered
already
as matrix models for {\it open} strings, as the logarithmic term has the effect
of
generating boundaries in the worldsheet \cite{6}.
In these open string models the couplings in front of the logarithm
and in its argument are fed in by hand, rather than being determined
by the original closed string action.

The simplest non-trivial action that can be concocted with a logarithmic
term, namely a linear plus logarithmic action, has been studied
quite extensively in the context of calculating the virtual Euler
characteristics for the moduli space of Riemann surfaces \cite{9}
as well as for being an interesting specimen of a matrix model from the string
theory point of view \cite{10,11}. If we use the notation of
Distler and Vafa \cite{11},
the partition function we wish to consider is
\begin{equation}
Z = \int d^{N^2} \Phi \exp N  t  \, tr \left[ \log ( 1  - \Phi ) + \Phi \right]
\end{equation}
where the integration is taken over positive definite $M = 1 - \Phi$ to avoid
problems arising from the branch cut in the logarithm. Penner showed that
the free energy $F = \log Z$ for this model could be expanded as
\begin{equation}
F =  \sum_{g,n} N^{2 - 2 g} t^{2 - 2 g - n } \chi_{g,n}
\label{e02}
\end{equation}
where the coefficients $\chi_{g,n}$, the virtual Euler characteristics of
the moduli space of
Riemann surfaces of genus $g$ with $n$ punctures, could be calculated
explicitly for all $g,n$
\begin{equation}
\chi_{g,n} = { (-1)^n ( 2 g - 3 + n ) ! ( 2 g - 1 ) \over ( 2 g ) ! n ! }
B_{2g}
\end{equation}
with the $B_{2g}$ being the Bernoulli numbers. Thus we know the expansion
of the free energy for {\it all} $g,n$ in this case.

We can make use of this fact to derive a ``renormalization group equation''
for the Penner model by simply differentiating equ.(\ref{e02}) directly,
instead of having to integrate out eigenvalues or rows and columns in a matrix.
If we reintroduce the customary factor of $1/N^2$
in front of $F$ and write
\begin{equation}
F_g = \sum_{n=1}^{\infty} N^{- 2 g} t^{2 - 2 g - n } \chi_{g,n}
\end{equation}
then it is a simple matter to show that
\begin{equation}
N {\partial F_g \over \partial N} - t {\partial F_g \over \partial t}  + 2 F_g
=
\sum_{n=1}^{\infty} n N^{- 2 g } t^{2 - 2 g - n}  \chi_{g,n}
\label{e03}
\end{equation}
where we have written explicitly the limits on the $n$ summation.
If we do not divide by $N^2$ the $2F_g$ drops out.
The right hand side of equ.(\ref{e03})
is almost
equal to $\partial F_g / \partial t$
which can be written,
after a shift in the summation variable $n \rightarrow n + 1$,
\begin{equation}
{\partial F_g \over \partial t} = \sum_{n=2}^{\infty}  N^{-2g} t^{2 - 2 g - n}
{(-1)^n ( 2 g - 3
+ n )! ( 2 g - 1) \over ( 2 g )! (n - 1) ! } B_{2g}.
\end{equation}
Only the $n=1$ term in the summation is missing,
so if we put this back by hand
we have, with the proviso $g>1$,
\begin{equation}
\sum_{n=1}^{\infty} N^{-2 g } t^{2 - 2 g - n}
n \chi_{g,n} = { \partial F_g \over \partial t} - N^{-2 g
} t^{1 - 2 g } {B_{2 g } \over 2 g }
\end{equation}
so
\begin{equation}
N {\partial F_g \over \partial N} - ( t + 1) {\partial F_g \over \partial t} +
2 F_g
= - N^{- 2 g} t^{1 - 2 g} {B_{2 g } \over 2 g }
\label{e04}
\end{equation}
which is a
renormalization group equation of the form in \cite{5}.
The critical coupling $t_c$ of the model is
determined by the zero of the beta function, $\beta(t_c)=0$.
As $\beta(t) = t + 1 $ by inspection of equ.(\ref{e04}),
this gives $t_c = -1$, which is the
correct result. Similarly, we should have
\begin{equation}
\gamma_1 = { 2 \over \beta'( t_c)}
\end{equation}
which trivially gives $\gamma_1 = 2$, thus recovering the
known fact that the Penner model
is a $c=1$ matrix model.
It is not surprising that we have obtained the exact values
for $t_c$ and $\gamma_1$, as we started with an exact expression for the
free energy, what is remarkable is that
we have not had to
include any terms that are nonlinear
in $\partial F_g / \partial t$ to write down the renormalization group
equation,
in direct contrast to the matrix models without logarithmic terms that were
considered in \cite{5}. In this respect the behaviour of the Penner
model is more like that of the vector models considered in \cite{4},
where the renormalization group equation is linear, even after using
reparameterization invariance to eliminate unwanted couplings.

So far, so good but we have not yet considered the $g=0,1$
cases and the presence of logarithmic corrections for these. Generically,
we would
expect to find a double zero in the beta function when
logarithmic corrections are
present. We are safe for $g>1$ as we know such corrections are absent, so the
renormalization group equation we have written down is at least consistent.
It turns out that the corrections are lurking in the lower
limits of the $n$ summations we have
carried out to evaluate $F_g$. In order to make sense of the sums
for $g=0,1$
Distler and Vafa regularized by
considering
the three punctured sphere
to kill the $SL(2,C)$ invariance and the once punctured torus
to kill the translational invariance
and then integrating to get the free energies without punctures.
In the matrix model we differentiate with respect to the
renormalized cosmological constant $\mu$, defined by
\begin{equation}
t = - 1 + {\mu \over N}
\end{equation}
to generate punctures.
If we reinstate the factor of $N^2$ in $F$ to
compare directly with the results of \cite{11}
and differentiate the expression for $F_0$
three times w.r.t. $\mu$, we find
\begin{equation}
F_0^{(3)} = - {1 / N} \sum_{n=1}^{\infty} ( 1 - \mu / N)^{- (n + 1)},
\end{equation}
where the superscript denotes the number of derivatives.
The factor of $1/N$ comes from the $N^2$ in the definition of $F_0$
multiplying
$1/N^3$ from the differentiations.
Performing the sum gives $1/ \mu$ as $N \rightarrow \infty$ and the three
integrations to get $F_0$ from $F_0^{(3)}$ generate the logarithmic
corrections.
Similarly, differentiating $F_1$ once gives
\begin{equation}
F_1^{(1)} =  {1 \over 12 N}  \sum_{n=1}^{\infty} ( 1 - \mu / N)^{- (n + 1)}
\end{equation}
which sums to $- 1/ 12 \mu$ as $N \rightarrow \infty$.
A single integration to get back to $F_1$ again produces the requisite
logarithm.

The renormalization group equation is phrased in terms of the bare
coupling $t$ rather than $\mu$, but we can still carry out the
summations in $F_0^{(3)}$ and $F_1^{(1)}$
in terms of $t$. Taking the torus first we find
\begin{equation}
F_1^{(1)} =  {1 \over 12} { 1  \over t ( 1 + t) }
\end{equation}
where the superscripts now denote differentiations w.r.t.
$t$.
By direct differentiation this satisfies
\begin{equation}
(t +1) {\partial F_1^{(1)} \over \partial t} + F_1^{(1)} = - { 1 \over 12 t^2}.
\end{equation}
If we differentiate our renormalization group equation w.r.t. $t$
(remembering there is no $2F_g$ term,
and that the factor of $N$ on the right hand side
is now $N^{2-2g}$ as we have retained
the overall factor of $N^2$ in $F$)
we get the same equation, so the results
are consistent.

For the three punctured sphere we have
\begin{equation}
F_0^{(3)} = - { N^2 } { 1  \over t ( 1 + t) }
\end{equation}
which, again by direct differentiation, satisfies
\begin{equation}
(t +1) {\partial F_0^{(3)} \over \partial t} + F_0^{(3)} =   {N^2 \over t^2}.
\end{equation}
If we differentiate our renormalization group equation three times
w.r.t. $t$,
use $N \partial F_0^{(3)} / \partial N = 2 F_0^{(3)}$
and set $g=0$ on the right hand side after performing the
differentiations  (as this is ill-defined otherwise)
we get precisely this equation.
Integrating $F_0^{(3)}$ once also gives the
correct value for the genus zero susceptibility
$\log(1+t) - \log(t)$.
We have thus shown that, although there is no double zero for $\beta(t)$, the
regularizations necessary for the sphere and the torus generate the
logarithmic terms in the free energy and the regulated free energies
$F_0^{(3)},F_1^{(1)}$
satisfy differentiated versions of the renormalization group equation.

We have in a sense cheated in this paper - because the form of the
free energy is exactly known we have not had to do any renormalization in order
to get our ``renormalization group equation''. If we look schematically
at a proper renormalization group calculation in the style of \cite{5}
where a matrix eigenvalue is integrated out, it is possible to see that
all the new induced terms are of the form of the original action up
to shifts in the coupling, so there is no need to introduce new couplings
and then use reparameterization invariance to gobble them up again.
It is rather remarkable that a complicated potential
with interactions of arbitrary order should give such a simple
beta function and not induce any further couplings under renormalization.
This may be because all of the possible vertices are already
included in the theory.
There could be some hints for attempts at finding a geometrical interpretation
for the matrix model renormalization group equation
for more general actions in all of this. We have found
for the Penner model
that a change in $N$ induces a flow in the coupling constant for
punctures in the surface, which is the only coupling in the theory
for this model.
The generation of an induced logarithmic term for other matrix models
suggests that the renormalization group flow
of puncture couplings might play an important role
here too, as put forward in the speculations at the end of \cite{11a},
which considered the baby universe structure of 2d quantum gravity.
The importance of punctures was also stressed in the
heuristic arguments of \cite{11b}, which suggested that
the $c=1$ barrier was the result of a condensation of ``spikes''
(ie punctures) in the Liouville theory.
It should be noted, however, that the change of scale in the matrix model
renormalization group is in the space of the matrix indices \cite{12}, so
a direct correspondence with real space renormalization
and finite size scaling ideas is not obvious.

D. Johnston was supported at Orsay by an EEC Human Capital and Mobility
fellowship
and an Alliance grant. A. Krzywicki is thanked for many discussions on the
utility (or otherwise!) of the Penner model and the matrix model
renormalization group.

\end{document}